\documentclass[prd,aps,a4,twocolumn,superscriptaddress,preprintnumbers,nofootinbib]{revtex4-1}
\usepackage{pslatex}
\usepackage[pdftex]{graphicx}
\usepackage{psfrag}
\usepackage{epsfig}
\usepackage{color}
\usepackage{cancel}
\usepackage{slashed}
\usepackage{amssymb}
\usepackage{amsmath}
\usepackage{hyperref}
\usepackage{enumerate}
\usepackage{multirow}
\usepackage{natbib}
\usepackage[normalem]{ulem}
\begin{document}
\title{Light vector mediators facing XENON1T data}%
\author{D. Aristizabal Sierra}%
\email{daristizabal@ulg.ac.be}%
\affiliation{Universidad T\'ecnica
  Federico Santa Mar\'{i}a - Departamento de F\'{i}sica\\
  Casilla 110-V, Avda. Espa\~na 1680, Valpara\'{i}so, Chile}%
\affiliation{IFPA, Dep. AGO, Universit\'e de Li\`ege, Bat B5, Sart
  Tilman B-4000 Li\`ege 1, Belgium}%
\author{V. De Romeri}%
\email{deromeri@ific.uv.es}%
\affiliation{Institut de F\'{i}sica Corpuscular CSIC/Universitat de Val\`{e}ncia, Parc Cient\'ific de Paterna\\
 C/ Catedr\'atico Jos\'e Beltr\'an, 2 E-46980 Paterna (Valencia) - Spain}%
\author{L. J. Flores}%
\email{luisjf89@fisica.unam.mx}%
\affiliation{Instituto de F\'isica, Universidad Nacional Aut\'onoma de M\'exico, A.P. 20-364, Ciudad de M\'exico 01000, M\'exico.}%
\author{D. K. Papoulias}%
\email{d.papoulias@uoi.gr}%
\affiliation{Department of Physics, University of Ioannina
GR-45110 Ioannina, Greece}%
\begin{abstract}
  Recently the XENON1T collaboration has released new results on
  searches for new physics in low-energy electronic recoils. The data
  shows an excess over background in the low-energy tail, particularly
  pronounced at about $2-3$ keV. With an exposure of $0.65$
  tonne-year, large detection efficiency and energy resolution, the
  detector is sensitive as well to solar neutrino backgrounds, with
  the most prominent contribution given by $pp$ neutrinos. We
  investigate whether such signal can be explained in terms of new
  neutrino interactions with leptons mediated by a light vector
  particle.  We find that the excess is consistent with this
  interpretation for vector masses below $\lesssim 0.1$ MeV. The
  region of parameter space probed by the XENON1T data is competitive
  with constraints from laboratory experiments, in particular GEMMA,
  Borexino and TEXONO.  However we point out a severe tension with
  astrophysical bounds and cosmological observations.
 
\end{abstract}
\maketitle
\section{Introduction}
\label{sec:intro}
Dark matter (DM) direct detection experiments have entered the era of
ton-size active volumes, and will keep going in that direction in
their search for DM signals
\cite{Aprile:2015uzo,Aalbers:2016jon,Aalseth:2017fik,Akerib:2018lyp,Schumann:2015cpa}. Combined
with high sensitivities at low energy thresholds as well as low and
fairly well-understood backgrounds, these experiments offer
opportunities in the search for DM signals which cover large classes
of DM physics models. Conventional searches using nuclear recoil
energy measurements allow searches of DM in the GeV-TeV range, while
electron recoil measurements provide a tool for sub-GeV DM and other
well-motivated degrees of freedom such as axion-like particles (ALPs)
and/or dark photons \cite{DiLuzio:2020wdo,Essig:2013lka}. Being
sensitive to irreducible solar neutrino backgrounds, they will enable
as well a better understanding of solar neutrino fluxes
\cite{Newstead:2018muu,Newstead:2020fie} and potentially new
directions in the search for new physics in the neutrino sector
\cite{Cerdeno:2016sfi,Dutta:2017nht,AristizabalSierra:2017joc,Gonzalez-Garcia:2018dep}.

XENON1T is a dual-phase liquid xenon time projection chamber with a
one-tonne active target \cite{Aprile:2015uzo}. The detector conceived
for WIMP DM searches in regions above $\sim 6\,$GeV can be used as
well for searches of ALPs, dark photons and neutrino properties,
thanks to the low energy thresholds and background rates. 
Given its dual-phase character, prompt scintillation and
delayed luminescence signals---$S1$ and $S2$---can be well
measured. Identification of electron and nuclear recoils can be done
through $S2/S1$ ratios, and thus provide a tool for particle
identification (e.g. neutron-induced nuclear recoils from
$\beta$-induced electron recoils). Recently XENON1T has released data
taken from February 2017 to February 2018, in which signals above
background-induced electron recoil events were searched for
\cite{Aprile:2020tmw}. The collaboration has reported an excess below
$7\,\text{keV}$ with a prominent feature towards
$2-3\,\text{keV}$. Using this data, three new physics scenarios were explored as
possible explanations to the signal: the solar axion model,
neutrino magnetic moment and bosonic dark matter. The finding shows that the resulting
$90\%\,$CL parameter space regions within which the excess can be
accounted for are disfavored by astrophysical arguments
\cite{Diaz:2019kim,Giannotti:2017hny,Corsico:2014mpa}. The
collaboration has as well tested this result against possible
background from tritium $\beta$ decays. In this case the statistical
significance of the new physics hypotheses is substantially
diminished.

Although the background hypothesis cannot be discarded, one can as well entertain the possibility that the excess is
driven by new physics. Indeed, seemingly, this has been the approach
the collaboration has adopted. 
If one is to adopt such approach as well, there is a lesson
one should take from the findings the collaboration has reported:
\textit{Whatever the nature of the new physics is, it should be able
  to produce localized spectral distortions}. Above $7\,$keV the data
is rather well described by the background, e.g. in the range
$25-50\,$keV the data points are beautifully accounted for by the
radio activity of $^{83m}$Kr \cite{Aprile:2020tmw}.

Possible scenarios of new physics are those in which electron recoils are modified by
the coupling of new degrees of freedom to electrons. The new degrees
of freedom could e.g. involve DM or particles from a dark sector
\cite{Takahashi:2020bpq,Kannike:2020agf,Alonso-Alvarez:2020cdv,Fornal:2020npv,Smirnov:2020zwf}. Another possibility, which
goes along the lines of the neutrino magnetic moment, is an
interaction that locally enhances the elastic scattering
neutrino-electron cross section \cite{1802141}. That is actually what
the neutrino magnetic moment does, it adds to the electroweak neutral
and charged current neutrino-electron cross section, dominating the
scattering process at low recoil energies. Possibilities include
neutrino non-standard interactions (NSI) as well as neutrino
generalized interactions (NGI)
\cite{Bolanos:2008km,Farzan:2017xzy,Coloma:2019mbs,AristizabalSierra:2018eqm,Khan:2019jvr},
both with electrons. In the effective
limit---$m_\text{Med}^2\gg q^2$ ($q$ being the exchanged
momentum)---these interactions will produce overall enhancements or
depletions (depletions only if the new interaction is driven by vector
boson exchange) of the SM expectation. Thus, given the typical
exchanged momentum, spectral distortions can be generated only by
light mediators, of which in this paper we consider the vector case.

The paper is organized as follows. In
Sec. \ref{sec:LVM} we discuss the relevant interactions, the
neutrino-electron differential cross section and the neutrino backgrounds
at XENON1T. In Sec. \ref{sec:param-space-analysis} we present our results: the event
rates, the phenomenological constraints on light vector states and a statistical analysis of the
parameter space. We summarize in
Sec. \ref{sec:conclusions}.
\section{Light vector mediator scenarios and 
  solar neutrino background}
\label{sec:LVM}

The interactions that we consider can be understood as a consequence
of a larger complete theory that we do not specify. They could be e.g. the
result of an extended gauge group $U_\text{B-L}$ or $U(1)_X$
\cite{Campos:2017dgc,Lindner:2018kjo}. 
For the purpose of this paper what matters is the presence of a new light vector mediator,
coupled to neutrinos and electrons, although one could in principle include the other charged leptons too. Along the same lines one could as well consider a kinetic mixing term,
between the hypercharge field of the SM and the new vector. For simplicity---however---we set
the tree level coupling to zero, bearing in mind that it then will be
generated radiatively and hence it will be suppressed.

The elastic neutrino-electron cross section involves the SM charged- and
neutral-current contributions as well as a new neutral-current
piece induced by the light vector. The cross section has been calculated in several
papers
\cite{Cerdeno:2016sfi,Ballett:2019xoj,Dent:2016wcr,Bolanos:2008km,Lindner:2018kjo}. Here we use the expression derived in Ref. \cite{Lindner:2018kjo}
which applies in $U(1)_X$ models in which neutrino and electron
charges $Q_X$ are dictated by anomaly cancellation conditions. It
reads
\begin{equation}
  \label{eq:x-sec}
  \frac{d\sigma}{dE_r}
  =\frac{m_eG_F^2}{4\pi}
  \left[
      g_2^2 
    + g_1^2\left(1 - \frac{E_r}{E_\nu}\right)^2
    - g_1g_2\frac{m_eE_r}{E_\nu^2}
  \right]\ ,
\end{equation}
where the couplings $g_{1,2}$ include both the SM and new physics
components as follows
\begin{equation}
  \label{eq:g1_and_g2_couplings}
  g_{1,2} = 
  g_{1,2}^\text{SM} 
  + 
  a_{1,2} + \frac{b_{1,2}}{G_F(2 m_e E_r + m_V^2)}\ .
\end{equation}

In the expression above, $G_F$ stands for the Fermi coupling constant, $E_r$ refers to the electron recoil energy, $E_\nu$ denotes the incoming neutrino energy, $m_e$ is the electron mass and 
$m_V$ the mass of the new vector mediator.
In the limit of suppressed 
mass mixing between the neutral vectors, the SM pieces
are given by their standard forms in terms of the weak-mixing angle $s_W^2$, as
\begin{align}
  \label{eq:SM-couplings}
  g_1^\text{SM}=-2\sqrt{2}s_W^2\ ,\qquad
  g_2^\text{SM}=\sqrt{2}(1 - 2s^2_W) - 2\sqrt{2}\ , 
\end{align}
with the second term in $g_2^\text{SM}$ being present only for electron
neutrinos. The SM limit is then given by
\begin{equation}
  \label{eq:x-sec-SM-limit}
  \frac{d\sigma_\text{SM}}{dE_r}=
  \left .\frac{d\sigma}{dE_r}\right|_{a_i=b_i=0}\ .
\end{equation}
Full expressions for the $a_i$ and $b_i$ parameters are given in Ref. \cite{Lindner:2018kjo}.
In the limits assumed in this paper (suppressed kinetic and mass mixing), they have a rather simple form
\begin{equation}
  \label{eq:a_i_b_i_couplings_zero_mixing}
  a_1=a_2=0\ ,\quad 
  b_1=-\frac{1}{4}Q_\nu^L Q_\ell^Rg_V^2\ ,\quad 
  b_2=-\frac{1}{4}Q_\nu^L Q_\ell^Lg_V^2\ ,
\end{equation}
where $g_V$ is the coupling associated to the new vector boson and
possible charge choices are determined by anomaly cancellation,
as in Ref. \cite{Campos:2017dgc}. Out of the possible
choices, $Q_\nu^L=Q_\ell^R=Q_\ell^L=-1$ corresponds to the well known
$U_\text{B-L}$ case. In the following we stick to this scenario.
From Eq. (\ref{eq:g1_and_g2_couplings}), one can see
that a spectral feature in the electron recoil events can be generated by the $q^2=-2m_eE_r$ dependence, 
as far as the vector is not decoupled. 
This is the limit we are interested in. 

\begin{figure}
  \centering
  \includegraphics[scale=0.9]{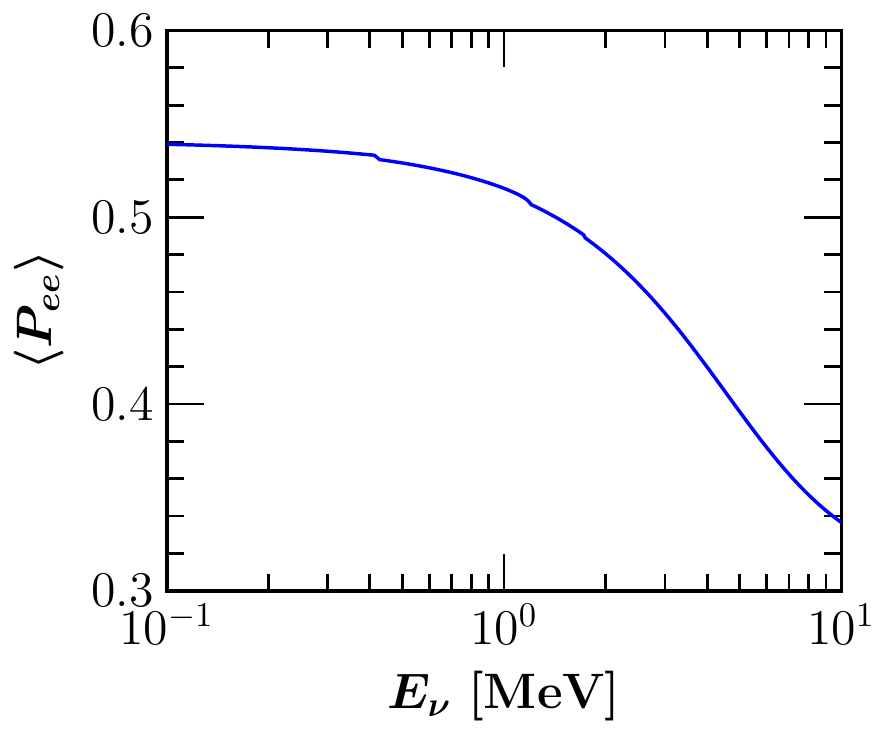}
  \caption{Averaged survival probability $\langle P_{ee}\rangle$ versus
    neutrino energy calculated in the two-flavor approximation with
    neutrino production distribution functions and neutrino fluxes
    ($pp$ and CNO) as given in the BS05 Standard Solar Model
    \cite{Bahcall:2004pz}. We have fixed the neutrino oscillation
    parameters according to their best fit point values
    \cite{deSalas:2017kay}.}
  \label{fig:probability}
\end{figure}
Having introduced the notation, we then move on to the determination of
the \textit{morphology} of the neutrino background at XENON1T. With a
0.65 tonne-year exposure, the expected number of solar neutrino
electron recoil events is $220.7\pm 6.6$ \cite{Aprile:2020tmw}. This
background can be obtained by integrating the following differential rate~\cite{AristizabalSierra:2017joc}
\begin{align}
  \label{eq:recoil-spectrum}
  \frac{dR}{dE_r}=\varepsilon\,N_T\sum_\alpha
  \int_{E_\nu^\text{min}}^{E_\nu^\text{max}}
    \frac{d\Phi_\alpha}{dE_\nu}\,
  \left[P_{ee}\frac{d\sigma_e}{dE_r}\right .
    \left . + (1-P_{ee})\frac{d\sigma_f}{dE_r}
  \right]\,dE_\nu\ ,
\end{align}
with $\alpha$ running over all the neutrino-related subprocesses of
the solar $pp$ and CNO chains: $pp$, $^8$B, $hep$, two $^7$Be and
$pep$ lines, $^{13}$N, $^{15}$O and $^{17}$F. Here $\varepsilon$
refers to the exposure in tonne-year,
$N_T=(Z_\text{Xe}/m_\text{molar})N_A$ to the number of target
electrons per tonne of material and $d\Phi_\alpha/dE_\nu$ to neutrino
flux in $\text{cm}^{-2}\text{year}^{-1}\text{MeV}^{-1}$ units. For the
solar neutrino fluxes we take the predictions of the BS05 Standard
Solar Model \cite{Bahcall:2004pz}.
\begin{figure}
  \centering
  \includegraphics[scale=0.9]{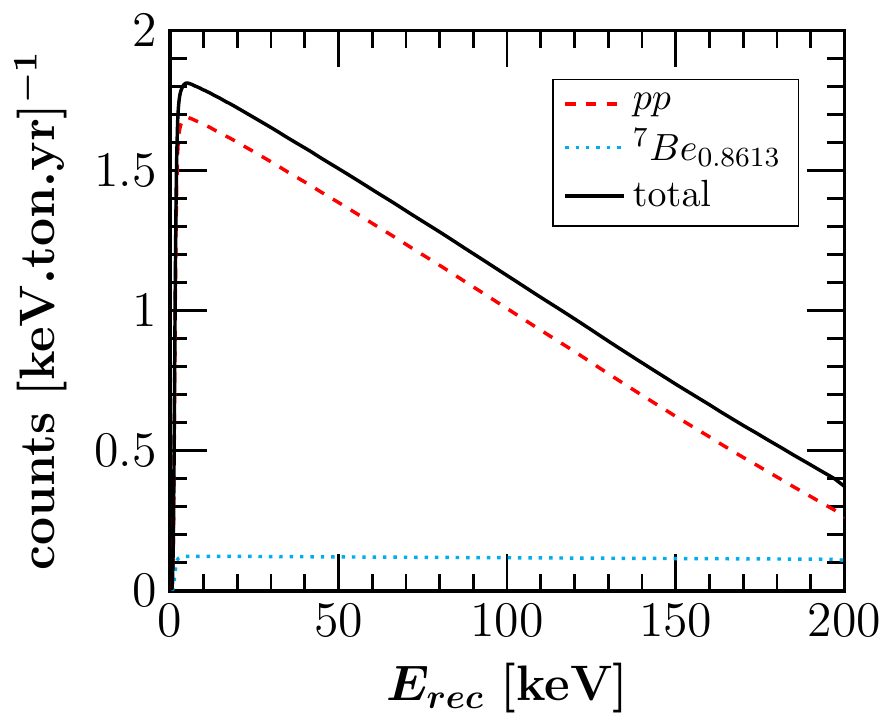}
  \caption{Expected number of neutrino-electron scattering events per
    tonne-year-keV induced by solar neutrinos. Most events are
    generated by the continuous $pp$ flux. $E_\text{rec}$ refers to
    reconstructed energy, see Sec. \ref{sec:param-space-analysis} for
    details.}
  \label{fig:sm-event-rate}
\end{figure}

Expression (\ref{eq:recoil-spectrum}) assumes the two-flavor
approximation, a fairly accurate limit given that
$\Delta m_{12}^2/\Delta m_{13}^2\ll 1$. In this limit one neutrino
eigenstate is mainly $\nu_e$ with flavor contamination suppressed by
the reactor mixing angle, while the other---labeled $f$---is a
superposition of $\nu_\mu$ and $\nu_\tau$ with the admixture
determined by the atmospheric mixing angle. The survival probability
proceeds from an average over the neutrino trajectory and weighted by solar
neutrino production distributions determined by the BS05 Standard
Solar Model. For its calculation we have proceeded as described in
Refs. \cite{Gonzalez-Garcia:2013usa,AristizabalSierra:2017joc}.
Figure \ref{fig:probability} shows our result derived for neutrino
oscillation parameters fixed according to the best fit point value
obtained from global neutrino oscillation data analysis:
$\Delta m_{12}^2=7.55\times 10^{-5}\,\text{eV}^2$,
$\sin^2\theta_{12}=0.32$ and $\sin^2\theta_{13}=0.0216$
\cite{deSalas:2017kay}.

The differential cross section in the first term of Eq.~(\ref{eq:recoil-spectrum}) is given by
Eq. (\ref{eq:x-sec-SM-limit}), the one in the second term as well but
without including the second term in $g_2^\text{SM}$ in
Eq. (\ref{eq:SM-couplings}). The lower integration limit is related to the
recoil energy through
\begin{equation}
  \label{eq:enumin}
  E_\nu^\text{min}=\frac{1}{2}\left(E_r + \sqrt{E_r^2 + 2E_rm_e}\right)\ ,
\end{equation}
while for $E^\text{max}_\nu$ we take the kinematic end points of each
of the neutrino fluxes.  Although the sum in
(\ref{eq:recoil-spectrum}) covers all neutrino emission processes,
given the recoil window, we find that 
the $pp$ continuous spectrum alone accounts for almost all the solar neutrino background.
This is somehow expected given the low energy threshold
achieved by the detector, $1\,$keV, and the size of the different
components of the neutrino flux. Figure~\ref{fig:sm-event-rate} shows
the differential event rate calculated with
Eq. (\ref{eq:recoil-spectrum}). There one can see that $pp$
neutrinos dominate the signal all over $E_r$. Other contributions are
subdominant, including the $^7$Be line at 0.861 MeV (the second relevant contribution) 
and $^8$B which for CE$\nu$NS will be the dominant
source \cite{Strigari:2009bq}.
\section{Constraints and parameter space analysis}
\label{sec:param-space-analysis}
In what follows we assume that only $\nu_e$ couples to the light new vector boson
while $\nu_f$ is subject only to SM couplings. Including coupling to $\nu_f$ will not
change our conclusions qualitatively. Although the new interaction can
affect neutrino propagation in matter, here it is reasonable to consider only effects in
detection. Forward coherent scattering is responsible for matter
effects, which given the solar electron density are prominent
(resonantly enhanced) only for $^8$B neutrinos. Since the signal is
driven by the $pp$ flux, propagation effects can be safely
ignored. Under those well-justified assumptions, the second term in
Eq. (\ref{eq:recoil-spectrum}) is negligible. However, for completeness, we keep the
full expression in our calculation.

\begin{figure}
  \centering
  \includegraphics[scale=0.9]{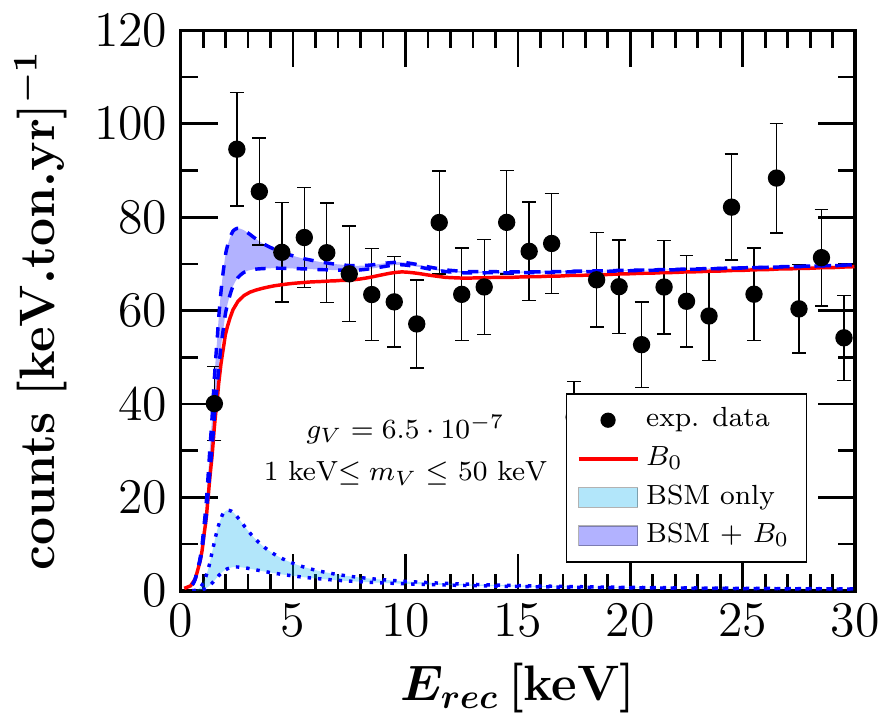}
  \caption{XENON1T data points along with the predicted background
    ($B_0$). The peaked signals at low energies are generated by a
    light vector mediator coupled to both electron-neutrinos and
    electrons and so contributing to the elastic scattering
    neutrino-electron cross section. The signals are generated using
    the benchmark points $m_V\subset [1,50]\;$keV and
    $g_V=6.5\times 10^{-7}$. $E_\text{rec}$ refers to reconstructed
    energy, see Sec. \ref{sec:param-space-analysis} for details.}
  \label{fig:neutrino-signals}
\end{figure}

Figure \ref{fig:neutrino-signals} shows
the effect of the new interaction on the
neutrino-electron differential event rate, along with the data points of XENON1T and the background $B_0$, assuming a fixed value of $g_V$ and scanning over $m_V$ 
in the range [1,50] keV (width of the light blue band). The signal peaks at low energies with
decreasing vector boson masses, for fixed coupling. 
This behavior is expected from the structure of the 
differential cross section. The third term in
Eq. (\ref{eq:g1_and_g2_couplings}) has a recoil energy dependence,
which becomes irrelevant for sufficiently large $m_V$. However, in the
limit $2m_eE_r\gg m_V^2$ that term behaves like $\sim E_r^{-1}$. Therefore, in
that regime, the $E_r$ dependence is 
basically the same of the $\nu-e$ neutrino magnetic moment
cross section~\cite{Vogel:1989iv,Bell:2006wi,Bell:2005kz}.

To statistically determine the regions favored/disfavored by XENON1T
data we define a simple spectral $\chi^2$ function as follows
\begin{equation}
  \label{eq:chiSq}
  \chi^2=\sum_{a=1}^{29}\frac{1}{\sigma_a^2}
  \left[
    \left(
      \frac{dR^{V+B_0}}{dE_{\rm rec}}
    \right)_a 
    -
    \left(
      \frac{dR^\text{Exp}}{dE_{\rm rec}}
    \right)_a
  \right]^2\ .
\end{equation}
Here $\sigma_a$ refers to statistical uncertainty per bin,
$E_{\rm rec}$ to reconstructed recoil energy and $B_0$ to
background. To compare with the experimental results from XENON1T we
have convolved the differential rate given in
Eq.~(\ref{eq:recoil-spectrum}) with a normalized Gaussian function
with an energy-dependent standard deviation defined as:
$\sigma/E_{\rm rec} = a/\sqrt{E_{\rm rec}} +b$,
$a=(31.71\pm0.65)\%$keV$^{1/2}$,
$b=(0.15\pm0.02)\%$~\cite{XENON:2019dti,Aprile:2020yad}. We further
apply the detector efficiency~\cite{Aprile:2020tmw}, after the
smearing.  $dR/E_\text{rec}$ in Eq. (\ref{eq:chiSq}) refers to the
quantity obtained that way. Note that given that the shift between the
nominal and reconstructed energies is always below $0.4\%$
\cite{Aprile:2020yad}, differences between results with or without
smearing are minor.

\begin{figure}
  \centering
  \includegraphics[scale=0.4]{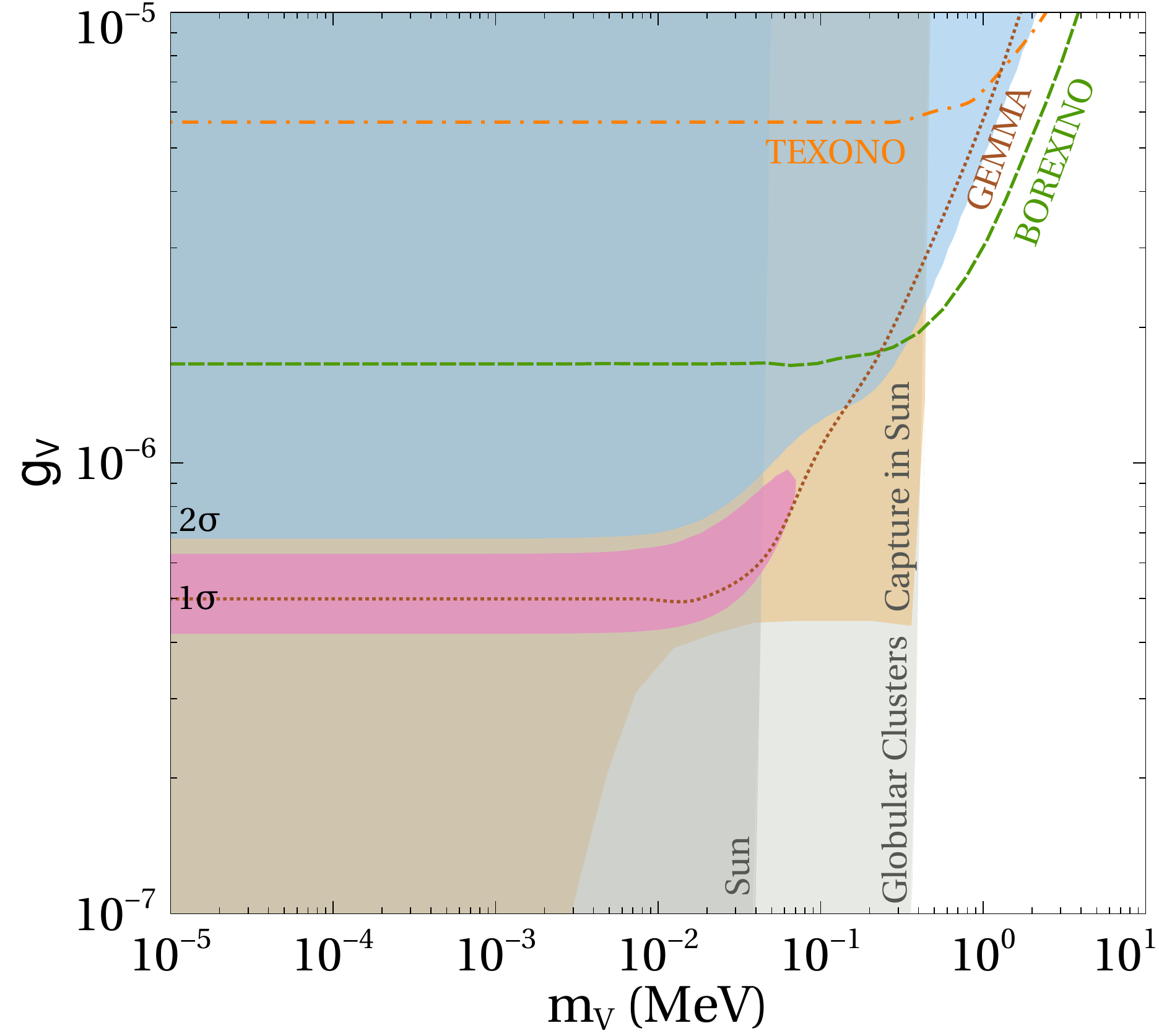}
  \caption{Allowed $1\sigma$ and excluded $2\sigma$ regions in the
    $m_V-g_V$ plane for light vector mediators. Along with the
    regions, laboratory limits from TEXONO \cite{Deniz:2009mu}, GEMMA
    \cite{Beda:2009kx,Beda:2010hk} and Borexino
    \cite{Agostini:2018uly,Bellini:2011rx} as well as those from
    astrophysics \cite{Grifols:1986fc,Grifols:1988fv,Chang:2018rso,Harnik:2012ni,Bilmis:2015lja}
   are shown as
    well.}
  \label{fig:excluded_allowed_regions}
\end{figure}

Following XENON1T analysis, we have added $B_0$ to the vector
contribution. The allowed $1\sigma$ region (pink) and $2\sigma$
excluded region (light blue region) resulting from our analysis are
shown in Fig. \ref{fig:excluded_allowed_regions}.  At the $1\sigma$
level, the allowed vector boson masses are always below
$\sim 800\;$keV with couplings that never exceed $6\times 10^{-7}$. In
this region the largest enhancements in the $2-7\,$keV energy range
are found.  As $m_V$ increases, the $E_r^{-1}$ behavior of the cross
section diminishes and the differential recoil spectrum flattens out
towards $B_0$.

At this point then the question is whether the $1\sigma$ allowed region is consistent with
existing bounds on light vector mediator scenarios, for example those in
\cite{Harnik:2012ni,Bilmis:2015lja,Bauer:2018onh,Cerdeno:2016sfi,Lindner:2018kjo} 
(some of them relevant also for CE$\nu$NS~\cite{AristizabalSierra:2019ufd,AristizabalSierra:2019ykk}). These
bounds can be separated in laboratory, astrophysical and cosmological
constraints. In the region of interest, the most stringent laboratory
limits are set by TEXONO, GEMMA and Borexino
\cite{Deniz:2009mu,Beda:2009kx,Beda:2010hk,Agostini:2018uly,Bellini:2011rx}, as shown in Fig.~\ref{fig:excluded_allowed_regions}.
XENON1T improves the constraints on light vector mediators for $m_V \lesssim 0.2$ MeV, compared to 
TEXONO and Borexino.
GEMMA limits are tighter but still leave unconstrained a fraction of the $1\sigma$ region. 
Overall, the regions within which XENON1T excess can be
accounted for are consistent with laboratory bounds.

Astrophysical and cosmological constraints are instead much more severe. 
The light vectors can be produced in environments like horizontal
branch stars and the Sun 
leading then to energy losses \cite{Grifols:1986fc,Grifols:1988fv,Dent:2012mx,Chang:2018rso}. The
presence of a vector neutrino coupling can affect the neutrino mean
free path in supernovae, eventually disrupting the neutrino
diffusion time \cite{Chang:2016ntp}. These arguments lead to stringent bounds
within the region of interest (see Fig.~\ref{fig:excluded_allowed_regions}).
In general, these limits might be evaded if the vector boson couples to light scalars that undergo condensation in the corresponding environment.
Under these conditions the vector mass
becomes environmental dependent and so its production is no longer
possible \cite{Nelson:2008tn}. 
Further relevant bounds come from cosmology. 
In the early Universe the vector boson can thermalize through
neutrino or electron annihilation or scattering
processes. This will alter the expansion history of the early Universe and eventually lead to a sizeable contribution to the effective number of neutrino species, $\Delta N_{\rm eff}$~\cite{Masso:1994ww,Ahlgren:2013wba,Kamada:2018zxi,Escudero:2019gzq,Dutta:2020jsy}. 
While the exact evaluation of these bounds will depend on the specific model and thermal history of the Universe, they appear to exclude the full XENON1T 1$\sigma$-region here derived.

\section{Conclusions}
\label{sec:conclusions}%
We have considered light vector mediator scenarios in the light of the
recent XENON1T data \cite{Aprile:2020tmw}. We have addressed the
question of whether these interactions can account for the spectral
distortion observed by the collaboration. Light vector mediators
generate spectral features, modifying the recoil energy dependence
of the differential cross section, which increases at low $E_r$ for sufficiently small vector
boson masses. 
We have performed a statistical analysis taking into account
the complete set of solar neutrino fluxes and the neutrino survival probability from oscillations.
We have shown that XENON1T bounds are competitive with those from other laboratory experiments.
Astrophysical and cosmological observations place instead more severe constraints, 
which potentially exclude the regions of parameter space which can explain the XENON1T 
excess.
With increasing exposures, multi-ton LXe detectors will keep
on testing other parameter space regions of these scenarios.

\vspace{1cm}
\noindent%
\textbf{Note added in proof}\\[2mm]
While completing this work, Ref. \cite{1802141} appeared in the arXiv
database. In addition to light vector mediators, this paper considered
as well light scalar mediators. Our findings are consistent with those
this reference has reported.
\section*{Acknowledgments}
DAS is supported by the grant ``Unraveling new physics in the
high-intensity and high-energy frontiers'', Fondecyt No 1171136.
VDR acknowledges financial support by the SEJI/2018/033 grant, funded by Generalitat Valenciana and partial support by the Spanish grants FPA2017-90566-REDC (Red Consolider MultiDark), FPA2017-85216-P and PROMETEO/2018/165 (Generalitat Valenciana). 
LJF is supported by a postdoctoral CONACYT grant.
The work of DKP is co-financed
by Greece and the European Union (European Social Fund- ESF) through
the Operational Programme ``Human Resources Development, Education and
Lifelong Learning" in the context of the project ``Reinforcement of
Postdoctoral Researchers - 2nd Cycle" (MIS-5033021), implemented by
the State Scholarships Foundation (IKY).

\bibliography{references}
\end{document}